\documentclass{elsart}%
\usepackage{epsfig} \usepackage{amsmath}
\usepackage{graphicx}%
\usepackage{amsfonts}%
\usepackage{amssymb}
                                                                                                                                               
\begin{document}
\begin{frontmatter}
\title{Efficient numerical integrators for stochastic models}
\author[CCS]{G. De Fabritiis \corauthref{cor1}}
\author[FIS]{M. Serrano}
\author[FIS]{P. Espa\~{n}ol}
\author[CCS]{P.V.Coveney}
\address[CCS]{ Centre for Computational Science, Department of Chemistry,\\
 University College London, 20 Gordon street, WC1H 0AJ, London, UK }
\address[FIS]{Departamento de F\'{\i}sica Fundamental, UNED,\\
Apartado 60141, 28080 Madrid, Spain}
 \corauth[cor1]{Corresponding author. E-mail address: g.defabritiis@ucl.ac.uk}
\begin{abstract}
The efficient simulation of  models  defined in terms of
stochastic differential  equations (SDEs)  depends critically on an efficient  integration
scheme. 
In  this article, we investigate under which conditions  the  integration schemes for general SDEs 
 can be derived using the Trotter expansion. It follows that, in the stochastic case, 
some  care is required in splitting the stochastic generator.  
We test the Trotter  integrators  on  an  energy-conserving Brownian  model  
 and derive a new numerical scheme for dissipative particle dynamics. 
We find  that the stochastic Trotter scheme provides a mathematically correct and easy-to-use method
which should find wide applicability.
\end{abstract}
\end{frontmatter}
\section{Introduction}
The study of  mesoscopic particle models  such as Brownian dynamics (BD)\cite{allen87},
 Dissipative Particle
Dynamics  (DPD)  \cite{hoog92,espanol95},  Smoothed Dissipative Particle  Dynamics  (SDPD)
\cite{revenga03}  and the Voronoi  fluid particle  model \cite{flekkoy00,serrano01}
requires efficient  integration methods  that solve the  appropriate stochastic
equations of  motion. In the past  few years, several  authors have considered
improvements to the  basic stochastic Euler schemes normally  applied to these
systems  of equations, particularly  in the  context of  ``conventional'' DPD.
Groot     \&    Warren     \cite{groot97},     Pagonabarraga   {\it et al.}
\cite{pagonabarraga98} and Besold  \emph{et al.} \cite{besold00} have reported
various performance improvements to the  basic schemes through the use of more
sophisticated  deterministic solvers,  for  example those  that have
been  successfully employed for  \emph{deterministic} dynamical  systems \cite
{channell90} including molecular  dynamics (MD) simulations \cite{tuckerman92}, such
as the velocity and leapfrog Verlet algorithms.
These  traditional   deterministic  integrators  provide  significant
improvements on  the basic Euler solver albeit,  being deterministic schemes,
their behaviour is completely uncontrolled from a theoretical point of view and their
order   of  convergence   is  not   clear.    In  fact,   these  solvers   
\emph{arbitrarily}  leave  out  terms  which  should  appear  in  a  correct
stochastic  expansion. 
More recently, alternative schemes have been devised
resulting from proper stochastic expansions \cite{shardlow03,nikunen03}, 
and even from a Monte Carlo-based approach \cite{lowe99,peters04}
where the fluctuations
are introduced via a thermostat (the deterministic dynamics
 is still dependent on the integrator). 

A general
method  for   deriving  deterministic integrators is based  on   the  Trotter  expansion
\cite{allen87,trotter59}.   For  Hamiltonian   systems,  these  schemes  preserve  the
symplectic structure of the dynamics  and conserve  the dynamical  invariants, ensuring  that the
long time behaviour is correctly captured.  In fact, if  a dynamical
invariant $I$  exists then the  discrete dynamics conserves exactly  a virtual
invariant $I^{\ast}$  which is bound to $I$  up to second order  in $\Delta t$
\cite{channell90}. 
 An important feature  of  mesoscopic models  is that they
often recover a  symplectic dynamics in some limit,  an  example being the DPD
model for vanishing friction coefficient.   It may be important to account for
this \emph{quasi-symplectic} property of the SDEs in the integration scheme by
assuring  that   in  the  same  limit   the  scheme  is   symplectic  as  well
\cite{mannella04}.

Recently, a first order stochastic generalisation of the Trotter expansion has
been rigorously proved \cite{tessitore01,kuhn01}. In fact, 
for specific stochastic equations there exist schemes up to weak fourth order
\cite{forbert00} or schemes corrected to reproduce more accurately 
the equilibrium distribution function \cite{mannella04}. 
The situation is less clear for a general SDE (such as Eq. (\ref{trotter_sde}) in Section \ref{section2}), 
for which the application of the Trotter formula was overlooked in the literature, thereby
generating some confusion in terms of how the Trotter formula 
can be used to split the stochastic equations.
It is therefore useful to investigate the applicability 
of the Trotter formula in the most general case. 
This is of direct relevance for mesoscopic models
which usually involve very large systems of SDEs.

The Trotter formula has been applied  to 
 devise efficient integrators for several specific mesoscopic models but often 
its use is limited to splitting the propagator into  several terms which 
are then integrated using standard numerical schemes. This approach would correctly produce 
the order of accuracy expected for the dynamics but potentially would affect adversely the conservation of the dynamical 
invariants or even detailed balance.
Examples include  a numerical scheme suggested by a straightforward application of the 
Trotter rule to the Voronoi fluid particle model equations \cite{defabritiis03} 
which leads to time steps that are two orders of magnitude larger  than the standard 
Euler scheme.   
In the context of the conventional DPD model, Shardlow \cite{shardlow03,nikunen03}  presented  a new
scheme,  which splits the stochastic  and deterministic parts
following the  Trotter rule, and then integrates the fluctuation-dissipation generators  using  the 
Bruenger \emph{et al.} scheme  \cite{bruenger84} tailored onto the DPD equations.
For Brownian dynamics, Ricci \& Ciccotti \cite{ricci03} derived  a numerical 
integrator based on the Trotter expansion which integrates  the propagators by using the Suzuki formula 
\cite{suzuki} to transform the time-ordered 
exponential solution of the Brownian dynamics equations into more tractable simple exponentials.

\section{Stochastic Trotter schemes}\label{section2}
Let     us    consider    first     a    deterministic     dynamical    system
$\mathbf{\dot{x}}(t)=\mathcal{L}[\mathbf{x}]$.   The formal  solution  of this
system  is  $\mathbf{x}(t)=  \sum_{p=0}^{\infty}\frac{1}{p!}(\mathcal{L}t)^{p}
[\mathbf{x}](\mathbf{x}_{0})  ( =e^{\mathcal{L}t}[\mathbf{x}](\mathbf{x}_{0}) ) $
as  can  be shown  from  the Taylor  expansion  around  the initial  condition
$\mathbf{x}_{0}$.   In general, the  operator can  be decomposed  into simpler
operators of  the form $\mathcal{L}=\sum_i^M  \mathcal{L}_i$. The Trotter formula (Strang \cite{strang68})
provides a straightforward approximation to the time propagator
\begin{equation}
e^{\mathcal{L}t}=\left( \prod_{i=M}^{1}e^{\mathcal{L}_{i}\frac{\Delta t}{2}
}\prod_{j=1}^{M}e^{\mathcal{L}_{j}\frac{\Delta t}{2}}\right) ^{P}+O(\Delta
t^{3})   \label{dettrotter}
\end{equation}
where $t=\Delta tP$, $P$ is the number  of time steps each of size $\Delta t$, and the ordering of the
$i,j$   indices    is   important.    In   the   case    that   two   operators
$\mathcal{A}$,$\mathcal{B}$    commute,    i.e.    $[\mathcal{A},\mathcal{B}]=
\mathcal{AB}-\mathcal{BA}=0,$ then  the approximate Trotter  formula is indeed
exact      because      the      equations      $e^{\mathcal{A}+\mathcal{B}}=
e^{\mathcal{A}}e^{\mathcal{B}} =e^{\mathcal{B}}e^{\mathcal{A}}$ are valid.  Because 
 the Trotter formula   decomposes the  dynamics over the
time interval $t$ into $P$ steps, it  provides a discrete algorithm for the solution of
the dynamics of the system.   Well known examples of the deterministic Trotter
expansion  are velocity  and position  Verlet schemes  for  molecular dynamics
simulations \cite{allen87}.

\begin{figure}
\begin{center}
\centerline{ \psfig{figure=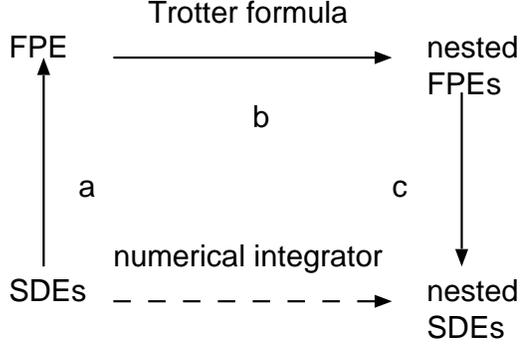,width=7cm} }
\end{center}
\caption{Diagram showing the derivation of SDE integrators by
  using the Trotter formula. Step (a) is the standard transformation from
  SDE  to  FPE  formalism  \cite{kloeden92}.  Step (b)  decouples  the  FPE  
  using the  deterministic  Trotter
  formula. Step (c) applies the  reverse transformation from each decoupled FPE to the
  corresponding  SDE in the order given in (b).  }
\label{diagram}
\end{figure}

In the stochastic case,
we define a $d$ dimensional stochastic process $
\mathbf{x}_{t}=(x_{t}^{1},...,x_{t}^{d})$ with associated stochastic
differential equation (SDE) in the It\^{o} interpretation
\begin{equation}
dx_{t}^{k}=a^{k}(\mathbf{x}_{t})dt+\sum_{j=1}^{m}b^{kj}(\mathbf{x}
_{t})dW_{t}^{j}   
\label{trotter_sde}
\end{equation}
where $a^{k}(\mathbf{x}_{t})$ is the drift vector, $b^{kj}(\mathbf{x}_{t})$ is
the  diffusion matrix  ($d$ variables,  $m$ Wieners)  and $dW_{t}^{j}$   the
vector of independent increments of  the $j$-th Wiener process.  
 The mathematically
equivalent  Fokker-Planck   equation  (FPE) of  Eq.    (\ref{trotter_sde})  for  the
probability density $\rho(\mathbf{x},t)$ is
\begin{equation}
{\partial_t}\rho=\mathcal{F}[\rho]   
\label{fokker}
\end{equation}
where     $\mathcal{F}[\rho]=-\sum_{k}\frac{\partial}{\partial    x^{k}}\left(
  a^{k}\rho\right)           +\frac{1}{2}\sum_{k,l}\frac{\partial^{2}}{\partial
  x^{k}\partial    x^{l}}\left(    d^{kl}\rho\right)$    and    $d^{kl}=\sum
  _{j}b^{kj}b^{lj}$ is the diffusion matrix. 

Following the  diagram depicted  in Fig. (\ref{diagram}), 
we  translate the  starting stochastic equation (\ref{trotter_sde}) into the
corresponding Fokker-Planck equation (\ref{fokker}) 
which has  formal solution     $\rho(\mathbf{x},t)=e^{\mathcal{F} t}[\rho](\rho_{0})$. 
The deterministic  Trotter formula (\ref{dettrotter})  can be applied  to this
formal  solution   by  generally  splitting   the  operator  $\mathcal{F}=\sum
_{i}\mathcal{F}_{i}$.   Furthermore, if  $\mathcal{F}_{i}$ is  a Fokker-Planck
operator  itself,  this picture  of  evolving  the  probability density  using the
Trotter formula  has a  counterpart at the  level of  the SDE which would allow us to devise a numerical integrator.
However, not  all  decompositions
$\mathcal{F}_{i}$ have  Fokker-Planck form and therefore an associated SDE.
 We then 
proceed by progressively splitting the terms in the starting SDE, i.e the drift vector $a^k$ and
the matrix $b^{kj}$, to verify Fokker-Planck  form.

The drift terms  do not present  any special
problem: that is  any splitting  of the vector 
\begin{equation}
 a^k=\sum_{\alpha} a^k_{\alpha},
\end{equation}
 produces  Fokker-Plank  drift-like  terms which can be easily integrated 
as with any standard  ordinary differential equation (ODE). 
The diffusion operator demands more care. The matrix $b^{kj}$ can
be  split  into columns  such  as  to give  several  systems  of single  noise
equations, $b_{\alpha}^{kj}=b^{kj}\delta_{\alpha,j}$  which are different from zero 
only in the  column   corresponding  to   noise  $\alpha=j$. By substituting $b^{kj}=\sum_{\alpha}b_{\alpha}^{kj}$ 
into the diffusive matrix $d^{kl}  
=\sum_{\alpha,\beta}\sum_{j}b_{\alpha}^{kj}b_{\beta}^{lj}$ we obtain 
\begin{equation}
d^{kl}=\sum_{\alpha}\sum_{j}b_{\alpha}^{kj}b_{\alpha}^{lj}
\end{equation}
 which is  split  into  several  diffusive  operators,  because  $\mathbf{b}_{\alpha}\cdot
\mathbf{b}_{\beta}^{t}=0,\forall\alpha\neq\beta$,   i.e.    the   correlations
between different diffusive dynamics are  zero. In this procedure, we decouple
the  diffusive dynamics  in terms  of  the subdynamics  corresponding to  each
independent Wiener  process.

We are still left to integrate $m$ single noise SDEs. 
We can try to decompose further each system of single noise SDEs into separate scalar SDEs.
For each noise $j$, we set  $ b_{\alpha}^{kj}={  b}^{kj}\delta_{\alpha,k}$ such that 
substituting in $d^{kl}$ we have
\begin{equation}
 d^{kl} =
 \sum_{\alpha,\beta}\sum_{j}b_{\alpha}^{kj}b_{\beta}^{lj},
\label{problem}
\end{equation}
which cannot be reduced to  Fokker-Planck form for all terms.  This
means that we cannot split variables over terms of the same noise to derive
the integrator. In fact, in order to
apply the diagram of Fig. (\ref{diagram}) and in particular step (c), we need to
have all the terms in Fokker-Planck form to derive the corresponding SDEs.
In principle, one could also try to separate the diffusion matrix
 $d^{kl}$ itself into several simpler  matrices $d^{kl}=\sum_{\alpha}d_{
\alpha}^{kl}$ provided that each  matrix $d_{\alpha}^{kl}$ is positive definite,
but  then  the  non-unique  square-roots  of the  matrices $d_{\alpha}^{kl}$  have to be computed
in  order to recover the SDEs.  Practically, this is very difficult in general.

Finally, we must be able to compute the solution
of the SDE corresponding to the $i$ term $\mathcal{F}_i$ in order to write down the integration scheme.
This is possible for simple SDEs, otherwise we can take advantage of the splitting 
between the drift and diffusion generators. 
The analytical solution of SDEs with zero drift is conveniently calculated in the Stratonovich 
interpretation for the stochastic integral (for a reference on Stratonovich integrals see \cite{kloeden92}). 
In fact,  the standard rules of ordinary calculus apply and the SDEs are effectively integrated like  ordinary 
differential equations by formally considering $dW$ as $dt$.
An It\^{o} SDE like Eq. (\ref{trotter_sde}) is transformed into the equivalent 
 Stratonovich form with the usual rules for the   drift  
\begin{equation}
 \underline{a}^{k}=a^{k}-\frac{1}{2}\sum_{j=1}^{m}\underline{L}
^{j}b^{kj}
\end{equation}
 where   $\underline{L}^{j}
=\sum_{h=1}^{d}b^{h,j}\frac{\partial}{\partial  x^{h}}$  and  the
noise                   term       is interpreted            accordingly                  as
$dx_{t}^{k}=\underline{a}^{k}(\mathbf{x}_{t})dt+\sum_{j=1}^{m}b^{kj}(
\mathbf{x}_{t})\circ  dW_{t}^{j}$ (see \cite{kloeden92}). 

As  the Trotter formula approximates the dynamics (\ref{fokker}) of the probability 
distribution $\rho$ up to  second order in time, we  expect that at the SDE level the accuracy
of  the   method  is  weak  second-order  \cite{kloeden92}, i.e. moments are accurate to second order.  
Effectively, the proposed decomposition at the FPE level  allows us to reduce the 
time-ordered exponential solution of SDE (\ref{trotter_sde})
in terms of simple exponentials  up to second order provided that the generators for the same  noise are not 
split.

\section{An energy-conserving Brownian model}
The oldest model for a stochastic system is the Langevin equation for a
Brownian particle. 
In the one dimensional case, the SDE governing the velocity
of the particle is $dv = -v dt + (2T)^{1/2}dW$ where we have selected units in
which the mass of the particle and friction coefficient are unity and $T$ is the
dimensionless bath temperature. This equation predicts an exponential decay of
the velocity and, consequently, of the kinetic energy of the Brownian
particle which  goes into the fluid surrounding the particle.
For illustrative purposes, we can construct an energy-conserving model in
which we include the energy $\epsilon$ of the fluid system, a Lagrangian reference system and a conservative force. 
We use the dimensionless equations in Stratonovich form
\begin{align}
dr & = v dt\nonumber\\
dv & =F(r)dt -vdt+(2\alpha\epsilon)^{1/2}\circ dW_{t},  \nonumber \\
d\epsilon & =v^{2}dt-(2\alpha\epsilon)^{1/2}v\circ dW_{t},
\label{browniansde}
\end{align}
where $F=-\frac{\partial V(r)}{\partial r}$ is the conservative force and $\alpha$  is a dimensionless heat  capacity of the fluid. 
 The above SDEs
have as a dynamical invariant the total energy $E=E_0=V(r)+\frac{v^{2}}{2}+\epsilon$. 
Generalisations of the SDEs (\ref{browniansde}) to higher dimensions and multiple particles 
are indeed fundamental building-blocks  of several mesoscopic models.

In practice, it is not necessary to move to a Fokker-Planck description to derive the integration scheme.
The derivation in section (\ref{section2}) shows that we can simply apply the Trotter formula (\ref{dettrotter}) over the generators 
of the  SDEs (\ref{browniansde}) provided that we do not split the stochastic generator for the same noise.
The SDEs (\ref{browniansde})  is written in the form $d{\bf x}_t = \mathcal{L}[{\bf x}]dt+ \mathcal{S}[{\bf x}]\circ d{\bf W}_t$, where
 ${\bf x}=(r,v,\epsilon)$ and 
 the deterministic and stochastic generators are respectively
$\mathcal{L} =  \mathcal{L}_1 + \mathcal{L}_2 + \mathcal{L}_3 +  \mathcal{L}_4$ and 
 $\mathcal{S}=\mathcal{S}_1+\mathcal{S}_2$,    
\begin{eqnarray}
\mathcal{L}_1 &=& v\partial/\partial r; 
~~\mathcal{L}_2 = F\partial/\partial v; 
~~\mathcal{L}_3 =-v\partial/\partial v ;
~~\mathcal{L}_4 = v^{2} \partial/\partial \epsilon ;\nonumber \\
\mathcal{S}_1 &=& (2\alpha\epsilon)^{1/2} \partial/\partial v ;  
~~\mathcal{S}_2= -(2\alpha\epsilon)^{1/2}v  \partial/\partial \epsilon.  
\end{eqnarray}
The generators $\mathcal{S}_1$ and $\mathcal{S}_2$ cannot be split and integrated independently using the Trotter formula
because they refer to the same noise. However, the solution for  $\mathcal{S}$ can be directly computed  by applying 
standard calculus on the system of two equations $d{\bf x}_t=\mathcal{S}[{\bf x}]\circ d{\bf W}_t$; 
the solution is given by
\begin{equation}
e^{\mathcal{S} \Delta W_{\Delta t}}[{\bf x}]:  
\begin{array}[l]{l}
 \epsilon \rightarrow C \cos\left (\sqrt \alpha \Delta W_{\Delta t} + \arccos(\sqrt{\epsilon/C})\right )^2,\\
 v \rightarrow sign(v) \sqrt{2C} \sin \left(\sqrt \alpha \Delta W_{\Delta t} + \arccos(\sqrt{\epsilon/C})\right ), 
\end{array}
\end{equation}
where  $sign(x)=1$ if $x \geq 0$ and $sign(x)=-1$ if $x < 0$.    
Both variables are updated starting from the same initial values 
and  $C=\epsilon +v^2/2$  is computed before the update.
The deterministic generators are easily integrated
\begin{eqnarray}
e^{\mathcal{L}_1\Delta t}[{\bf x}]&: &r \rightarrow r+v \Delta t;
~~e^{\mathcal{L}_2\Delta t}[{\bf x}]: v \rightarrow v+F \Delta t,\nonumber\\
e^{\mathcal{L}_3\Delta t}[{\bf x}]&: &v \rightarrow v\exp(-\Delta t);
~~e^{\mathcal{L}_4\Delta t}[{\bf x}]: \epsilon \rightarrow\epsilon+v^{2}\Delta t.
\end{eqnarray}
The solutions of these differential equations can be nested following any given order to obtain different  integration schemes.
A possible numerical scheme is 
\begin{equation}
e^{\mathcal{S}\Delta W_{\Delta t/2}}
e^{\mathcal{L}_4\frac{\Delta t}{2}}
e^{\mathcal{L}_3\frac{\Delta t}{2}}
e^{\mathcal{L}_2\frac{\Delta t}{2}}
e^{\mathcal{L}_1\Delta t}
e^{\mathcal{L}_2\frac{\Delta t}{2}}
e^{\mathcal{L}_3\frac{\Delta t}{2}}
e^{\mathcal{L}_4\frac{\Delta t}{2}}
e^{\mathcal{S} \Delta W'_{\Delta t/2}},\label{brownianscheme}
\end{equation}
where  $\Delta W'_{\Delta t/2}$ and $\Delta W_{\Delta t/2}$ are two random numbers drawn 
from a zero mean normal distribution with standard deviation $\sqrt{\Delta t/2}$.
We note that the stochastic propagator of this scheme conserves energy exactly (for any time step size), therefore the conservation 
of energy depends only on the approximation introduced in the deterministic part. 
 
As already stated, it is not possible to decompose the stochastic generator $ \mathcal{S}$ 
into two independent stochastic scalar equations
using the Trotter formula. 
Unfortunately, this approach is what  would follow  if one was to apply naively the Trotter formula to  SDE (\ref{browniansde}). 
The resulting scheme would not be second order and would conserve energy poorly. For instance, this is the case for the scheme  
\begin{equation}
e^{\mathcal{S}_1\Delta W_{\Delta t/2}}
e^{\mathcal{S}_2\Delta W_{\Delta t/2}}
e^{\mathcal{L}_4\frac{\Delta t}{2}}
e^{\mathcal{L}_3\frac{\Delta t}{2}}
e^{\mathcal{L}_2\frac{\Delta t}{2}}
e^{\mathcal{L}_1\Delta t}
e^{\mathcal{L}_2\frac{\Delta t}{2}}
e^{\mathcal{L}_3\frac{\Delta t}{2}}
e^{\mathcal{L}_4\frac{\Delta t}{2}}
e^{\mathcal{S}_2\Delta W'_{\Delta t/2}}
e^{\mathcal{S}_1 \Delta W'_{\Delta t/2}},
\label{brownianwrong}
\end{equation}
where the stochastic propagators are
\begin{eqnarray}
e^{\mathcal{S}_1 \Delta W_{\Delta t}}[{\bf x}]&:&
v \rightarrow v+\sqrt{2\alpha\epsilon}\Delta W_{\Delta t},\nonumber\\
e^{\mathcal{S}_2\Delta W_{\Delta t}}[{\bf x}]&:&
\epsilon \rightarrow(\sqrt{\epsilon}-\sqrt{2\alpha}v/2\Delta W_{\Delta t})^{2}.
\end{eqnarray}

Interesting, there is a possibility to apply a Trotter-like rule to devise second order weak integrators 
even for the decomposition  $\mathcal{S} = \mathcal{S}_1+\mathcal{S}_2$. To do this the noises have to be advanced 
by $\frac{\Delta W_{\Delta t}}{2} =(weak)  \Delta W_{\Delta t/4}$,
 where by $=(weak)$ we mean that  moments of both sides are equal  to second order. 
 Note that for the Trotter expansion it should be
 $\Delta W_{\Delta t/2}=W_{t+\Delta t/2}-W_{t}$. 
The scheme is written as  
\begin{equation}
e^{\mathcal{S}_1\Delta W_{\Delta t/4}}
e^{\mathcal{S}_2\Delta W_{\Delta t/4}}
e^{\mathcal{L}_4\frac{\Delta t}{2}}
e^{\mathcal{L}_3\frac{\Delta t}{2}}
e^{\mathcal{L}_2\frac{\Delta t}{2}}
e^{\mathcal{L}_1\Delta t}
e^{\mathcal{L}_2\frac{\Delta t}{2}}
e^{\mathcal{L}_3\frac{\Delta t}{2}}
e^{\mathcal{L}_4\frac{\Delta t}{2}}
e^{\mathcal{S}_2\Delta W_{\Delta t/4}}
e^{\mathcal{S}_1 \Delta W_{\Delta t/4}},
\label{brownianscheme2}
\end{equation}
where we use the same realization of the noise  $\Delta W_{\Delta t/4}$.
The second order weak convergence can be verified by a direct comparison with a second order stochastic expansion and 
intuitively understood by  formally considering  $\Delta W$ as $\Delta t$.
We stress that the resulting scheme does \emph{not} correspond to a stochastic Trotter expansion, 
but rather to a second order approximation of the  propagator.
This method provides a way to write an integration scheme even in cases where it is impractical
 to compute the  solution of  the generator $\mathcal{S}$ altogether. 
However,  wherever possible, this approach should be avoided or limited  to the smallest generator 
because the resulting integration scheme may loose important structural features of the dynamics  
(as in the example of SDEs (\ref{browniansde})). 

\begin{figure}
\begin{center}
  \psfig{figure=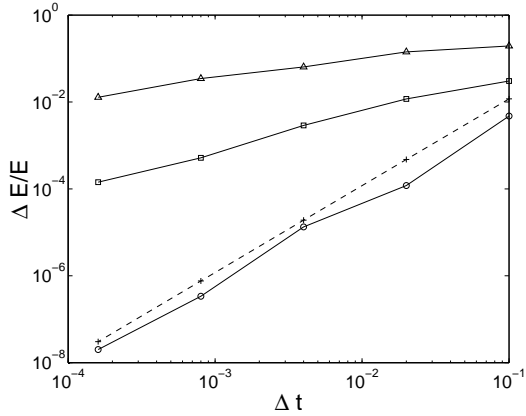,width=7cm}
\end{center}
\caption{
Average of the maximum relative error  of the energy for the SDE (\protect\ref{browniansde}) over 10 independent runs up to $t=1$ for the
  Trotter schemes (\ref{brownianscheme}) circles, (\ref{brownianscheme2}) squares and the incorrect Trotter scheme 
(\ref{brownianwrong}) triangles. 
The deterministic Trotter scheme for Eq. (\protect\ref{browniansde}) with $\alpha=0, v_0=1,\epsilon_0=1/2$ is plotted with dotted lines for reference.
 }
\label{brownian_fig}
\end{figure}

We validated numerically the integration schemes (\ref{brownianscheme}) and (\ref{brownianscheme2}) as well as  the incorrect one 
(\ref{brownianwrong}).
The simulations were run using the bistable potential $V(r)=\beta(r^4-2r^2)$ with 
 $\alpha=1$, $\beta=1$ and initial conditions  $r_0 = 0,~v_{0}=0$
and $\epsilon_{0}=1$. 
The average relative error for the total energy $\Delta E/E$ for different  time step lengths $\Delta t$ is shown in Fig.
\ref{brownian_fig}.  The error  is computed by averaging the maximum error reached by $t=1$ over
 10 independent runs.
The stochastic-Trotter scheme (\ref{brownianscheme}) conserves the energy 
with the same accuracy as the deterministic Trotter scheme (computed using $\alpha=0$). 
The  scheme (\ref{brownianscheme2}) is consistent with first order accuracy
 (it is second order for single time step error),
 while the incorrect  scheme (\ref{brownianwrong}) does not conserve energy with first order accuracy. 
Note that the order for the cumulative error is one less than the single time step error. 
Clearly, the  energy conservation performance of the Trotter scheme (\ref{brownianscheme}) 
 is a direct consequence of the exact integration of its stochastic component which is impossible to achieve by other  general schemes.

\section {A Trotter integration scheme for dissipative particle dynamics}

We now apply the stochastic Trotter expansion to the equations of dissipative particle dynamics.
The DPD model consists of a set of $N$ particles moving in continuous space. Each
particle $k$ is defined by its position $\mathbf{r}_{k}$ and its momentum $%
\mathbf{p}_{k}$ and mass $m$. The dynamics is specified by a set of
Langevin equations very similar to the molecular dynamics equations, but where
in addition to the conservative forces there are   dissipative 
and  fluctuating forces as well
\begin{eqnarray}
d\mathbf{{r}}_{k} &=&\mathbf{p}_{k}/m dt, \nonumber\\
d\mathbf{{p}}_{k} &=&\sum_{l\neq k}^{N} \mathbf{e}_{kl}\left [  a_{kl} F_c(r_{kl})
dt - \gamma/m \omega _{D}(r_{kl})(\mathbf{e}_{kl}\cdot
\mathbf{p}_{kl})dt + \sigma \omega
_{R}(r_{kl})dW^t_{kl}\right ],
\label{dpd}
\end{eqnarray}
where $F_c(r)$ is the conservative pair interaction force weighted by  positive and symmetric parameters  $a_{kl}$,
  $\mathbf{r}_{kl}={\bf r}_k - {\bf r}_l$
is the distance between the particle $k$ and particle $l$,   ${r}_{kl}$ its length 
and ${\bf e}_{kl}={\bf r}_{kl}/r_{kl}$.
The weight functions $\omega_D, \omega_R$ usually have finite range $r_c$
and are related by $\omega_D(r_{kl})=\omega_R^2 (r_{kl})$ in order to satisfy detailed balance. 
 This condition  ensures that the equilibrium state is  Gibbsian
 and sets the value of its temperature to $T_0=\frac{\sigma^2}{2\gamma
   k_B}$. A typical selection is $\omega _{R}(r_{kl})=\omega(r_{kl})$ with
\begin{equation}
\omega(r)=\left\{
\begin{array}{cc}
1-\frac{r}{r_{c}} & r<r_{c} \\
0 & r\geq r_{c}.
\end{array}
\right.
\end{equation}
The  conservative force ${ F}_c(r_{kl})=- \frac{\partial V(r_{kl})}{\partial
r_{k}} $ is usually chosen to be of the form ${ F}_c(r_{kl})=w(r_{kl})$.

The generator of DPD equations (\ref{dpd}) is  
$\mathcal{L}=\sum_k 
\mathcal{L}^k_r + \sum_{k,l\neq k}\left (\mathcal{D}^{kl}+\mathcal{S}^{kl} \right ) ,$
where 
\begin{eqnarray}
\mathcal{L}^{k}_r&=& \mathbf{p}_{k}/m \partial/\partial {\bf r}_k; 
~~\mathcal{S}^{kl}=\sigma \omega_{R}(r_{kl})\mathbf{e}_{kl} \partial/\partial {\bf p}_k; \nonumber\\
\mathcal{D}^{kl}&=&a_{kl} F_c(r_{kl}) \mathbf{e}_{kl} \partial/\partial {\bf p}_k
 -\gamma/m \omega _{D}(r_{kl})(\mathbf{e}_{kl}\cdot \mathbf{p}_{kl})\mathbf{e}_{kl} \partial/\partial {\bf p}_k.
\end{eqnarray} 
In the DPD model the momentum is conserved because the forces between
 interacting particles $k$ and $l$  satisfy Newton's third law.
 We  split the DPD equations in order to satisfy this requirement. 
The conservative and fluctuation-dissipation generators  for the pair interaction  
 $k$, $l$  give
\begin{equation}
d{\bf x}=(\mathcal{D}^{kl}+\mathcal{D}^{lk})[{\bf x}] dt+(\mathcal{S}^{kl}+\mathcal{S}^{lk})[{\bf x}]dW^t_{kl}
\end{equation}
where ${\bf x}=({\bf r}_1,{...},{\bf r}_N,{\bf p}_1,{...},{\bf p}_N)$.
The solution  is  computed  by noting that $d\mathbf{{p}}_{k}+d \mathbf{{p}}_{l}=0$ and 
 $d\mathbf{{p}}_{k}=\frac{1}{2}d\mathbf{{p}}_{kl}$ where $d\mathbf{{p}}_{kl}=d\mathbf{{p}}_{k}- d\mathbf{{p}}_{l}$.  
The equation for  $d\mathbf{{p}}_{kl}$ can be solved for the component of the radial direction because
from the form of the SDEs (\ref{dpd}) it follows that 
$d\mathbf{{p}}_{kl} =d(\mathbf{{p}}_{kl}\cdot\mathbf{e}_{kl}) \mathbf{e}_{kl}$. 
Let us call $p^e_{kl}=\mathbf{{p}}_{kl}\cdot\mathbf{e}_{kl}$; then we have an 
Ornstein-Uhlenbeck process 
\begin{equation} 
dp^e_{kl}= A dt - B p^e_{kl}dt + C dW^t_{kl},
\end{equation}
 where $A=2a_{kl}F_c(r_{kl})$, $B=2 \gamma/m \omega_D$ and $C=2 \sigma \omega_R$,
which has  analytical solution \cite{kloeden92}
\begin{equation}
p^e_{kl}(t)= e^{-B \Delta t}  p^e_{kl}(t_0) 
+A \int_{t_0}^{t}  e^{B (s-t)}ds
+C \int_{t_0}^{t} e^{B(s-t)}dW_s, 
\label{exact}
\end{equation} 
where $\Delta t = t-t_0$, $t_0$ being the initial time. 
The solution (\ref{exact}) of the Ornstein-Uhlenbeck process requires the generation of coloured noise based on a numerical scheme itself \cite{fox88}.
In fact, the stochastic process $p^e_{kl}(t)$ has stationary correlation function for $t,s \rightarrow \infty$ with finite $|t-s| 
$ given by
\begin{equation}
<p^e_{kl}(t)p^e_{kl}(s)>=\frac{A^2}{B^2}+\frac{C^2}{2B}\exp(-B|t-s|).
\end{equation}
A version of the method  to generate  coloured noise \cite{fox88} adapted to Eq. (\ref{exact}) results in the scheme 
 \begin{equation}
\Delta {p}^e_{kl} = \left (\mathbf{{p}}_{kl}\cdot{\bf e}_{kl} - \frac{a_{kl}F_c}{\frac{\gamma}{m} w_D}\right)
 \left(e^{-2\frac{\gamma}{m} \omega_D \Delta t}-1 \right ) 
	   + \frac{ \sigma \omega_{R} \sqrt{1 -  e^{-4 \gamma/m \omega _{D} \Delta t}}   }{\sqrt{ \gamma /m \omega_D}}\xi^{kl},  
\end{equation}
where $\xi^{kl}=\xi^{lk}$ are normal distributed with zero mean and variance one ($N(0,1)$) and $\Delta {p}^e_{kl}={p}^e_{kl}( t) - {p}^e_{kl}( t_0)$ .

The propagator  $\mathcal{K}^{kl}$   
for $\mathbf{p}_{k}$ and $\mathbf{p}_{l}$ is then given by 
\begin{equation}
\mathcal{K}^{kl}_{\Delta t}[{\bf x}]: 
(\mathbf{{p}}_{k},\mathbf{{p}}_{l}) \rightarrow
\left (\mathbf{{p}}_{k} + \frac{1}{2} \Delta {p}^e_{kl} {\bf e}_{kl},~~
 \mathbf{{p}}_{l} - \frac{1}{2} \Delta {p}^e_{kl} {\bf e}_{kl} \right).\label{scheme_nc}
\end{equation}
%
The remaining position update is given by 
\begin{equation}
e^{\mathcal{L}^k_r\Delta t}[{\bf x}]: {\bf r}_k \rightarrow {\bf r}_k+ {\bf p}_k/m \Delta t.
\label{schemev}
\end{equation}
We note that $\mathcal{L}^k_r$ commutes with $\mathcal{L}^l_r$, therefore we can use the exact formula 
$e^{\sum_k \mathcal{L}^k_r\Delta t}=\prod_{k=1}^{N}e^{\mathcal{L}^k_r\Delta t}$.

The DPD scheme is finally given by the following Trotter integrator
\begin{equation}
{\bf x}(t+\Delta t)\rightarrow  \prod_{k=1,l=1}^{N}
\mathcal{K}^{kl}_{\Delta t/2}
~~
\prod_{k=1}^{N}e^{\mathcal{L}^k_r\Delta t}
\prod_{k=N,l=N}^{1}
\mathcal{K}^{kl}_{\Delta t/2}~{\bf x}(t).
\label{dpdscheme}
\end{equation}

In practice the integration algorithm consists of the following steps: 
for the interaction pairs k,l update the momentum half timestep according to the propagator (\ref{scheme_nc}), 
where  $\xi^{kl}=\xi^{lk}$ are drawn
 from a normal distribution with zero mean and variance one;
iterate over particles $k$ updating the position according to  (\ref{schemev}); finally,
update  pairs k,l in reverse order again using the propagator (\ref{scheme_nc}) but
 with new noise $\xi'_{kl}$. 
This algorithm requires the calculation of the pair-list only once per iteration and has the same complexity as a simple  
 DPD velocity-Verlet scheme (DPD-VV \cite{groot97}).  

We test this integration scheme using the open-source code {\em mydpd} \cite{mydpd} written in simple C++ and implementing
 the DPD models described here with periodic boundary conditions.
 The simulations are run with $N=4000$ particles, $a_{kl}=25$, $\gamma=4.5 $, $\sigma=3$, $m=1$, $r_c=1$ 
in a three dimensional periodic box $(L,L,L)$ with $L=10$. These settings give a particle density $\rho=4$ 
and equilibrium temperature $k_B T=1$.
In our implementation, the computational cost of each scheme averaged over several iterations indicates that the Trotter 
scheme is 60\%  more costly than the simple DPD-VV but 10\% faster than the Shardlow S1 scheme (which costs almost twice than DPD-VV). 
The equilibrium temperature for the DPD-Trotter scheme of Eq. (\ref{dpdscheme}), 
 DPD-VV \cite{groot97} and Shardlow \cite{shardlow03} schemes is reported in Table \ref{tab:table}. 
The DPD-Trotter scheme recovers the equilibrium temperature 
 better than DPD-VV, and as accurately as Shardlow's scheme.
This difference  depends on the implicit scheme used by Shardlow for the integration of the pair interaction. 
In our case, we have  used an exact integration  Eq. (\ref{exact}) 
which, however, requires the generation of  coloured noise \cite{fox88} which is by itself
 a numerical scheme.
Considering the accuracy of the equilibrium temperature and the computational cost,  
both DPD-Trotter and Shardlow schemes are integrators of comparable performance for the DPD equations.  
A more detailed study of the equilibrium properties of the fluid is necessary to assess the accuracy 
in reproducing the equilibrium distribution and other statistical properties.

\begin{table}
\caption{\label{tab:table} 
Equilibrium temperature for the DPD-Trotter, Shardlow and DPD-VV schemes for different time steps. 
The average of the kinetic temperature  $<k_B T>$ is computed  over a simulation of duration $t=1000$. 
The standard deviation of the estimates, computed by block-averaging, is less than $\pm 5 \times 10^{-4}$.
}
\begin{center}
\begin{tabular}{cccc}
\\
\hline
\hline
$\Delta t $ & DPD-Trotter (scheme Eq. (\ref{dpdscheme})) & Shardlow \cite{shardlow03} & DPD-VV \cite{groot97}\\
\hline
0.05 &  1.0136 & 1.0138   & 1.0411 \\
0.02 &  1.0020 & 1.0018   & 1.0097 \\
0.01 &  1.0007 & 1.0005   & 1.0043 \\
\hline
\hline
\\
\end{tabular}
\end{center}
\end{table}

\section{Conclusions}
The stochastic Trotter schemes can provide efficient integrators for stochastic 
models with dynamical invariants  by    fully
taking into account  the underlying stochastic character. 
The stochastic Trotter formula can be applied to any model based on SDEs and should find wide applicability
provided that some care is used to  decouple the stochastic dynamics for the same noise.
These types of stochastic schemes offer the flexibility to easily tailor the integrator to the specific model, thereby 
 integrating exactly important parts of the dynamics.
This  stochastic Trotter scheme is a second order weak scheme, but, more important,  in our examples 
it provides  very good conservation  of the dynamical invariants.

{\bf Acknowledgements}

We thank G. Tessitore for useful comments. This work was partially supported by the SIMU Project, European Science
Foundation. GDF is supported by the EPSRC Integrative Biology project GR/S72023. 
M.S. and P.E. thank the Ministerio de Ciencia y Tecnolog\'{\i}a, Project BFM2001-0290. 
\bibliographystyle{elsart-num} \bibliography{../../gianni}

\begin{thebibliography}{10}
\expandafter\ifx\csname url\endcsname\relax
  \def\url#1{\texttt{#1}}\fi
\expandafter\ifx\csname urlprefix\endcsname\relax\def\urlprefix{URL }\fi

\bibitem{allen87}
{M.~P. Allen and D.~J. Tildesley}, Computer Simulations of Liquids, Oxford
  University Press, Oxford, 1987.

\bibitem{hoog92}
{P.~J.~Hoogergrugge and J.~M.~V.~A. Koelman}, Europhys. Lett. 19 (1992) 155.

\bibitem{espanol95}
P.~{E}spa{\~n}ol, P.~Warren, Europhys. Lett. 30 (1995) 191.

\bibitem{revenga03}
P.~Espa{\~n}ol, M.~Revenga, Phys. Rev. E 67 (2003) 026705.

\bibitem{flekkoy00}
E.~G. Flekk{\o}y, P.~V. Coveney, G.~{De Fabritiis}, Phys. Rev. E 62 (2000)
  2140.

\bibitem{serrano01}
M.~Serrano, P.~Espa{\~{n}}ol, Phys. Rev. E 64 (2001) 046115.

\bibitem{groot97}
R.~D. Groot, P.~B. Warren, J. Chem. Phys. 107 (1997) 4423.

\bibitem{pagonabarraga98}
I.~Pagonabarraga, M.~H.~J. Hagen, D.~Frenkel, Europhys. Lett. 42 (1998) 377.

\bibitem{besold00}
G.~Besold, I.~Vattualainen, M.~Karttunen, J.~M. Polson, Phys. Rev. E 62 (2000)
  R7611.

\bibitem{channell90}
P.~J. Channell, C.~Scovel, Nonlinearity 3 (1990) 231.

\bibitem{tuckerman92}
M.~Tuckerman, B.~J. Berne, J. Chem. Phys. 97 (1992) 1990.

\bibitem{shardlow03}
T.~Shardlow, SIAM J. Sci. Comput. 24 (2003) 1267.

\bibitem{nikunen03}
P.~Nikunen, M.~Karttunen, I.~Vattulainen, Comp. Phys. Comm. 153 (2003) 407.

\bibitem{lowe99}
C.~P. Lowe, Europhys. Lett. 47 (1999) 145.

\bibitem{peters04}
E.~A. J.~F. Peters, Europhys. Lett. 66 (2004) 311.

\bibitem{trotter59}
H.~F. Trotter, Proc. Amer. Math. Soc. 10 (1959) 545--551.

\bibitem{mannella04}
R.~Mannella, Phys. Rev. E 69 (2004) 041107.

\bibitem{tessitore01}
G.~Tessitore, J.~Zabczyk, Semigroup Forum 63 (2001) 115.

\bibitem{kuhn01}
F.~Kuhnemund, Bi-continuous semigroup on spaces with two-topologies: theory and
  applications, Ph.D. thesis, Eberhard Karls Universit{\"{a}}t T{\"{u}}bingen,
  Germany (2001).

\bibitem{forbert00}
H.~A. Forbert, S.~A. Chin, Phys. Rev. E 63 (2000) 016703.

\bibitem{defabritiis03}
G.~{De Fabritiis}, P.~V. Coveney, Comp. Phys. Comm. 153 (2003) 209.

\bibitem{bruenger84}
A.~Bruenger, C.~L. {Brooks~III}, M.~Karplus, Chem. Phys. Lett. 105 (1984) 495.

\bibitem{ricci03}
A.~Ricci, G.~Ciccotti, Mol. Phys. 101 (2003) 1927.

\bibitem{suzuki}
M.~Suzuki, Proc. Jpn. Acad. B 69 (1993) 161.

\bibitem{strang68}
G.~Strang, SIAM J. Numer. Anal. 5 (1968) 506.

\bibitem{kloeden92}
P.~E. Kloeden, E.~Platen, Numerical solution of stochastic differential
  equations, Springer-Verlag, Berlin, 1992.

\bibitem{fox88}
R.~F. Fox, I.~R. Gatland, R.~Roy, G.~Vemuri, Phys. Rev. A 38 (1988) 5938.

\bibitem{mydpd}
Available online at http://www.openmd.org/mydpd.

\end{thebibliography}
\end{document}